\documentstyle{article}
\begin{document}
\renewcommand{\baselinestretch} {1.5}
\large
\hyphenation{anti-ferro-mag-netic}

\begin{flushright}
{\large{\bf YAMANASHI-95-02\\ December 1995}}
\end{flushright}
\vskip 0.5in
\begin{center}
{\large{\bf Study on Ground States of Quantum Spin Systems \\
by Restructuring Method}}
\vskip 0.5in
   Tomo Munehisa and Yasuko Munehisa\\
\vskip 0.3in
   Faculty of Engineering, Yamanashi University\\
   Kofu, Yamanashi, 400 Japan\\
\vskip 0.7in
{\bf Abstract}
\end{center}

We present a new method to study the ground state of
quantum spin systems using the Monte Carlo techniques together
with restructured intermediate states which we proposed previously.

Our basic idea is to obtain coefficients in the expansion of the ground
state from measurements of the normalized frequencies of each state
on the boundary in the Trotter direction.
The restructuring method we have proposed makes it possible to
carry out simulations within available resources of computers.

For a concrete example we study ladder spin systems
with antiferromagnetic intra-chain nearest neighbor interactions,
whose ground states may be equivalent to that of the spin 1 chain.
For the ladder with ferromagnetic rungs we successfully show how
to determine the magnitudes and relative signs of the
coefficients in the expansion.
When the ladder has antiferromagnetic rungs
we examine White's claim that the diagonally situated pair of spins
are mostly triplet states in the ground state.
Our results strongly suggest that the system has the hidden order which
is constructed by the specially aligned triplet states of diagonal pairs.

\eject
\noindent {\bf Section 1 \ Introduction}

Recent development of experiments on the condensed matter
has presented us very interesting systems where strong quantum
effects will be realized.
Among them are the quantum spin systems in low dimensions.
Much theoretical work has been done to obtain qualitative and
quantitative properties of these quantum effects \cite{thesis}.
One powerful tool to numerically investigate them
is Monte Carlo approach using the Suzuki-Trotter formula \cite{st}.
In a previous paper \cite{mune}, where we proposed the restructuring
method, we pointed out that proper choice of complete sets which describe
the intermediate states in the Suzuki-Trotter formula is quite important
in numerical calculations. Then we applied our method to a quantum spin
chain system to show that by employing complete sets different from
the conventional ones we can obtain improved numerical results.
An essential point of our method is to rearrange spin states on
neighboring two sites into the singlet and the triplet states when
we construct the complete sets.

In this paper we discuss another merit of our method which
comes out in study of the system's ground state \cite{Hatano}.
In order to provide a concrete example
we apply the method to quantum spin systems on a simple ladder
\cite{Hida} which have attracted many people for two reasons.
One reason is that this model is an intermediate one between one spin
$1$ chain which has a finite energy gap (Haldane gap) and two spin
$1/2$ chains which are gapless.
Another reason is that the corresponding material exists and this
would show the high temperature superconductivity upon some doping.

An outline of our basic idea is given in the next section together with
the model we study as an example.
In section 3 we present numerical results and
the final section is devoted to discussions.

\vskip 0.5in
\noindent {\bf Section 2 \ Model and Basic idea}

The Hamiltonian we study is
\begin{eqnarray}
   \hat{H} = - {J \over 2}\sum_{i=1}^N
             [ \vec{\sigma}_{a,i} \vec{\sigma}_{a,i+1}
            +  \vec{\sigma}_{b,i} \vec{\sigma}_{b,i+1} ]
 -{J' \over 2} \sum_{i=1}^N \vec{\sigma}_{a,i}\vec{\sigma}_{b,i},
\label{eq:H}
\end{eqnarray}
where $\vec{\sigma}_{a(b),i}$ denotes the Pauli matrix sitting on the
$i$-th rung of the leg $a(b)$ and
$\vec{\sigma}_{a(b),N+1} \equiv \vec{\sigma}_{a(b),1}$,
$N$ being the total number of rungs.
Let us set the intra-chain coupling $J = -1$ throughout this paper
and investigate the model with both positive and negative inter-chain
couplings $J'$.

The basic idea is quite simple.
In the system's partition function $Z$ its ground state $\mid G \rangle$
dominates when the inverse temperature $\beta $ is sufficiently large,
\begin{eqnarray}
   Z = tr( e^{ -\beta \hat{H}} )
   \simeq e^{ -\beta E_G} tr(\mid G \rangle \langle G \mid )
    = e^{-\beta E_G} \sum_\alpha \langle \alpha \mid G \rangle
                                 \langle G \mid \alpha \rangle ,
\label{eq:Z}
\end{eqnarray}
where $E_G$ denotes the energy of the ground state and
$\lbrace \mid \alpha \rangle \rbrace$ is a complete set.
Accordingly we expect that the partition function $Z_{n_t}$ in the
Suzuki-Trotter formula with finite Trotter number $n_t$ would be
\begin{eqnarray}
\nonumber
 Z_{n_t} & = & \sum_{\alpha_1}\sum_{\alpha_2} \cdots \sum_{\alpha_{n_t}}
\langle \alpha_1 \mid e^{-\beta \hat{H}/n_t} \mid \alpha_2 \rangle
\cdots \langle \alpha_{n_t}\mid e^{-\beta \hat{H}/n_t} \mid \alpha_1
\rangle \\
 & \simeq & \sum_{\alpha_1}
\langle \alpha_1 \mid G \rangle e^{-\beta E_G}
\langle G \mid \alpha_1 \rangle
\label{eq:Zts}
\end{eqnarray}
for large $n_t$ and $\beta$.
Let us expand $\mid G \rangle$ by $\lbrace \mid \alpha \rangle \rbrace$,
\begin{eqnarray}
\mid G \rangle = \sum_\alpha c_\alpha \mid \alpha \rangle,
\label{eq:G}
\end{eqnarray}
where the coefficients $c_\alpha$'s should be real numbers
in our representation.

With this expansion $(\ref{eq:Zts})$ reads
\begin{eqnarray}
Z_{n_t} \simeq e^{-\beta E_G} \sum_{\alpha_1} c_{\alpha_1}^2
 = e^{-\beta E_G} \sum_{\alpha} c_{\alpha}^2.
\label{eq:Ztse}
\end{eqnarray}
This means that $c_\alpha^2$ should be the probability to find state
$\mid \alpha \rangle$ on the boundary in the Trotter direction.

Since the number of $\mid \alpha \rangle$'s, which totals to $2^{2N}$
on a ladder with $N$ rungs, is enormously large for large $N$ it is
technically important to choose a proper complete set for which many
of $c_\alpha$'s vanish.
In the conventional approach the complete sets $\lbrace \mid \alpha _n
\rangle \rbrace$ $(n=1, \ldots , n_t)$ in $(\ref{eq:Zts})$ are
constructed from eigenstates of $\sigma^z_{a(b),i}$ with eigenvalues
$ \pm 1$ denoted by $\pm _{a(b),i}$ or by $s_{a(b),i}$.
The complete sets should be therefore
\begin{eqnarray}
\mid \alpha _n \rangle = \mid s_{a,1}, s_{b,1}, s_{a,2}, s_{b,2}, \cdots
                              s_{a,N}, s_{b,N} \rangle.
\label{eq:Ac}
\end{eqnarray}
In our approach we rearrange states $s_{a,i}$ and $s_{b,i}$ into
a singlet state $ \mid \ominus_i \rangle $ and three triplet states
$ \mid 1_i \rangle$, $\mid \oplus_i \rangle$ and $\mid -1_i \rangle$,
\begin{eqnarray}
\nonumber
\mid \ominus _i\rangle & = & {1 \over \sqrt{2}} \left (
\mid +_{a,i},-_{b,i} \rangle - \mid +_{a,i},-_{b,i} \rangle \right ), \\
\nonumber
\mid 1_i \rangle & = & \mid +_{a,i}, +_{b,i} \rangle , \\
\nonumber
\mid \oplus _i \rangle & = & {1 \over \sqrt{2}} \left (
\mid +_{a,i},-_{b,i} \rangle + \mid +_{a,i},-_{b,i} \rangle \right ), \\
\mid -1_i \rangle & = & \mid -_{a,i}, -_{b,i} \rangle ,
\label{eq:Si}
\end{eqnarray}
which we denote by $S_i$.  The complete sets we employ are
\begin{eqnarray}
\mid \alpha _n \rangle = \mid S_1, S_2, \cdots S_N \rangle.
\label{eq:Ar}
\end{eqnarray}
We will see later that this choice effectively reduces the number of
non-zero $c_\alpha$.

What we can measure in simulations with $Z_{n_t}$ is not $c_\alpha$
itself but $c_\alpha^2$.
One may have a question if there is a way to know the relative sign of
each $c_\alpha$. The answer is yes in the quantum spin case. All we have
to do is to change the boundary condition in the Trotter direction from
the periodic one to a combination of an open boundary
$\mid \alpha \rangle$ and a fixed boundary $\mid \alpha_0 \rangle$.
The corresponding partition function $Z'_{n_t}$ is
\begin{eqnarray}
\nonumber
 Z'_{n_t} & = &
\sum_{\alpha} \sum_{\alpha_2} \cdots \sum_{\alpha_{n_t}}
\langle \alpha_0 \mid e^{-\beta \hat{H}/n_t} \mid \alpha_2 \rangle
\cdots
\langle \alpha_{n_t}\mid e^{-\beta \hat{H}/n_t} \mid \alpha \rangle
\\ & \simeq & \sum_{\alpha}
\langle \alpha_0 \mid G \rangle e^{-\beta E_G}
\langle G \mid \alpha \rangle
 = e^{-\beta E_G} \sum_{\alpha} c_{\alpha_0} c_{\alpha}
\label{eq:Zptse}
\end{eqnarray}
Thus we will obtain $c_\alpha / c_{\alpha_0} =
c_{\alpha_0} c_\alpha / c_{\alpha_0}^2$ in simulations with $Z'_{n_t}$.

With the normalization condition $\sum c_\alpha^2 =1$
it would seem that the results with $Z'_{n_t}$ are enough to determine
both the magnitude and the sign of the coefficients $c_\alpha$'s but,
as we will see in the next section, simulations with $Z_{n_t}$ are
quite useful to obtain reliable values of magnitude with small
statistical errors.



\vskip 0.5in
\noindent {\bf Section 3 \ Numerical results}

In this section we study a ladder spin system with the Hamiltonian
$(\ref{eq:H})$ using Monte Carlo techniques. Exact values on a small
ladder are also presented to confirm the Monte Carlo results.

First let us work with positive $J'$s, which mean ferromagnetic rungs.
As we stated in Section~2 we expect that by applying the restructuring
method we can do with less number of non-zero coefficients in the
expansion $(\ref{eq:G})$ than using the conventional complete sets.
In order to see this in a quantitative manner let us examine
distributions of $c_\alpha^2$ on an $N=4$ and $J'=1$ ladder
calculated by the exact diagonalization. With the conventional complete
sets $(\ref{eq:Ac})$ the number of non-zero $c_\alpha^2$'s is sixty-eight
in total, twenty of which are greater than $10^{-2}$ and
sum over these twenty $c_\alpha^2$'s amounts to $0.808$.
With the rearranged complete sets $(\ref{eq:Ar})$, on the other hand,
total number of non-zero coefficients reduces to thirty-eight and
among them $c_\alpha^2$'s that are greater than $10^{-2}$ are fifteen
in number and $0.843$ in their sum.

In addition, for the system under consideration
it is believed that the ground state is essentially equivalent
to that of spin 1 chain, which is made of triplet states
on sites and seems to have
the hidden order that demands the $\mid 1 \rangle$ and
$\mid -1 \rangle$ states should appear alternately along the chain
with arbitrary number of $\mid \oplus \rangle$'s between them \cite{Aff}.
The complete sets $(\ref{eq:Ar})$ are much more suitable to describe
such a ground state.
The exact results on the $N=4$ and $J'=1$ ladder stated above,
for example, show that all the fifteen states in $(\ref{eq:Ar})$
whose $c_\alpha^2$'s are greater than $10^{-2}$
are of this type while others are not.
For an $N=8$ ladder we do not carry out exact calculations any more,
but it is easy to see that we have
only 255 candidate states in our choice,
which can be sorted to 22 states if we take account of the translational
invariance due to the periodic boundary condition in the space direction.
Contributions of these particular states
on larger ladders will be examined in the following Monte Carlo study.

Now we present numerical results with the partition function
$(\ref{eq:Ztse})$ and the complete sets $(\ref{eq:Ar})$.
In this case we do not need to worry about the negative
sign problem.
We measure frequencies of $\mid \alpha \rangle$s' appearance
on the boundary in the Trotter direction as well as system's energy
\begin{eqnarray}
E = -{\partial \over {\partial \beta}} \ln Z \; .
\label{eq:E}
\end{eqnarray}
In Figure~1 we plot $-E$ per site on an $N=8$ and $J'=1$ ladder
for several values of $\beta$ as a function of $1/n_t^2$,
with Trotter numbers ranging $16 \leq n_t \leq 32$.
Since $n_t$ dependence of physical quantities are known to be $1/n_t^2$
we make a linear fit for data with each value of $\beta$ to obtain the
energy in the $n_t \rightarrow \infty $ limit.
We observe the extrapolated values almost coincide,
which indicates that these values of $\beta$ are large enough
to realize the ground state.
In Figure~2 we compare Monte Carlo results on normalized frequencies
measured on an $N=4$ ladder at $\beta=8$ with exact $c_{\alpha}^2$
numerically obtained by the exact diagonalization.
We see the agreement is satisfactory at this value of $\beta$.
Figure~3 shows sum of contributions of the states with the hidden
order on an $N=8$ ladder at $\beta =8$ for several values of $J'$
in the $n_t \rightarrow \infty$ limit.
We observe the sum, which indicates sum of $c_\alpha^2$ for these
state, increases toward unity as we change $J'$ from $0.5$ to $5$.
Data for values of $N=4, 12$ and $16$ with $J'=1$
are also plotted in Figure~3 to show the dependence on $N$.
Since the results for $N=8$ show no drastic change as a function of $J'$
our data support that this system and the spin 1 chain are equivalent
in properties of the ground state.
The results also indicate that the hidden order dominates if
$J'$ is quite large, while it is less clear for smaller values of $J'$.

Next we present results on the coefficient itself measured with the
partition function $(\ref{eq:Zptse})$ instead of $(\ref{eq:Ztse})$.
Note that in this case, because of the emergence of negatively signed
weights, we should measure a physical quantity $A$ in each thermally
equilibrated configuration together with its sign and obtain its
expectation value by
\begin{eqnarray}
\langle A \rangle = {A_{net} \over Z_{net}}
 = {{A_+ - A_-} \over {Z_+ - Z_-}},
\label{eq:A}
\end{eqnarray}
where $Z_+ (Z_-)$ is number of configurations with positive (negative)
weight and $A_+ (A_-)$ denotes the contribution from positively
(negatively) signed configurations.
Because of the cancellations the statistical error is larger for
smaller $r$ ratio, which is defined by \begin{eqnarray}
r \equiv  {{Z_+ - Z_-} \over {Z_+ + Z_-}}.
\label{eq:r}
\end{eqnarray}
Here we limit ourselves to an $N=4$ and $J'=1$ ladder and set the
fixed boundary in the Trotter direction $\mid \alpha_0 \rangle = \mid
\oplus ,\oplus ,\oplus ,\oplus \rangle $.
We then measure normalized frequencies of $\mid \alpha \rangle$s'
appearance on the open boundary in the Trotter direction, $f_\alpha$,
following to $(\ref{eq:A})$.
Since $f_\alpha$ should be proportional to $c_\alpha c_{\alpha_0}$
we obtain the value of $c_{\alpha_0}^2$ from
\begin{eqnarray}
\sum \left ({f_\alpha \over f_{\alpha_0}} \right ) ^2 =
\sum \left ({c_\alpha c_{\alpha_0} \over c_{\alpha_0}^2} \right ) ^2 =
\sum \left ({c_\alpha \over c_{\alpha_0}} \right ) ^2 =
{{\sum c_\alpha^2} \over c_{\alpha_0}^2} = {{1} \over {c_{\alpha_0}^2}},
\label{eq:ca0}
\end{eqnarray}
using the normalization condition $\sum c_\alpha^2 =1$.
We then calculate $c_\alpha$, not $c_\alpha^2$, by multiplying
$\sqrt {c_{\alpha_0}^2}$ to each $f_\alpha / f_{\alpha_0}$, where we
assume $c_{\alpha_0}$ is positive.
Figure~4 plots the results together with the exact $c_\alpha$'s.
We observe that the data provide relative signs of $c_\alpha$'s to
$c_{\alpha_0}$ completely correctly
in spite of rather poor value of the $r$ ratio (which is around 0.15)
because $Z_- (Z_+) = 0 $ for positive (negative) $c_\alpha$.
The statistical errors on the magnitude of $c_\alpha$ is, however,
much worse than the ones measured
with the partition function $(\ref{eq:Ztse})$.

Finally we study the system with antiferromagnetic rungs,
namely with negative $J'$ in $(\ref{eq:H})$, which
became much interesting recently because of White's conjecture \cite{Wh}.
In the limit of large $\mid J'\mid$ this model has
the ground state that is a cluster of singlet at rungs as well
as the excited state including a triplet state at some rung, with
which the system has a finite mass gap. Since many studies
\cite{many} suggest this mass gap survives for finite
$\mid J' \mid$, one is tempted to think the system might be equivalent to
the spin 1 chain. In his study based on the density renormalization group
equation, White claimed that it is the case. It is reported
that roughly $95 \%$ of the {\it diagonal}
pairs on a ladder are triplet states in the system's ground state.
Here we examine the model by employing the complete sets constructed
with eigenstates of the diagonal pair,
\begin{eqnarray}
\mid \alpha _n \rangle = \mid S'_1, S'_2, \cdots S'_N \rangle,
\label{eq:Ard}
\end{eqnarray}
where $S'_i$ is one of
\begin{eqnarray}
\nonumber
\mid \ominus' _i\rangle & = & {1 \over \sqrt{2}} \left (
\mid +_{a,i},-_{b,i+1} \rangle - \mid +_{a,i},-_{b,i+1} \rangle \right ), \\
\nonumber
\mid 1'_i \rangle & = & \mid +_{a,i}, +_{b,i+1} \rangle , \\
\nonumber
\mid \oplus' _i \rangle & = & {1 \over \sqrt{2}} \left (
\mid +_{a,i},-_{b,i+1} \rangle + \mid +_{a,i},-_{b,i+1} \rangle \right ), \\
\mid -1'_i \rangle & = & \mid -_{a,i}, -_{b,i+1} \rangle .
\label{eq:Sid}
\end{eqnarray}
Using these complete sets with the partition function (\ref{eq:Ztse}),
we measure sum of normalized frequencies of the states which have the
hidden order.
Unfortunately we find that serious cancellations
prevent us from obtaining statistically meaningful data
except for small ladder size $N$ or small values of $\beta$ shown
in Figure~5.
In the figure we see, however, that data for a $N=4$ ladder are enough
to indicate this ratio increases rapidly as $\beta$ grows and at
$\beta=5$ it is in good agreement with the exact value shown
by a dashed line. Rapid increments as a function of $\beta$
are also observed for $N=8$, $12$ and $16$ ladders
as far as we could measure.
So we expect similar saturations will take place for larger ladders
with large values of $\beta$.
We therefore conclude that the hidden order in this model is
numerically observed.

\vskip 0.5in
\noindent {\bf Section 4 \ Discussions}

In this paper we present a new method to study the ground state of
quantum spin systems which uses the Monte Carlo techniques applied
to the Suzuki-Trotter formula. Studying the ground state by the
Monte Carlo approach will help us to understand its physical meanings,
to make effective models and to easily measure some physical quantities.

Our basic idea is to observe the
states emerging on the boundary in the Trotter direction and
count number of appearances of each state so that we can obtain
the normalized frequencies which are related to the coefficients
in the expansion of the ground state.
This method is based on the restructuring method we proposed
in a previous paper, which would enable us to effectively
reduce the number of candidate states we should consider.
We would like to stress here that it is essential from technical
points of view to choose some proper complete sets, although
any complete sets will do us from theoretical points of view.

In order to make our discussions clear we study ladder spin systems.
For the ladder whose rungs are ferromagnetic we show it is
possible to describe its ground state using the Monte Carlo results
obtained in our method.
For the ladder with antiferromagnetic rungs our results support
White's report that most diagonal pairs on the
ladder are the triplet states.
Our data also strongly suggest that the hidden order exists.

Our method would also be applicable to other systems such as
spin systems on a square lattice.
In order to construct proper complete sets, however, one should have
physical insights to the system under investigation.

\eject

\noindent {\bf Figure Captions}

\noindent {\bf Figure \ 1} \\
System's energy $E$ as a function of the Trotter number $n_t$ on
an $N$=8 and $J'=1$ ladder.
The ordinate and the abscissa represent $-E$ per site and $1/ n_t^2$,
respectively.
Each datum is an average of twelve measurements with different random
number sequences, for each of which we generate 10000 configurations
for the thermalizaion and more 10000 configurations for the measurement.
Statistical error of the datum is calculated from the standard deviation
among these twelve runs.
Here we do not show statistical errors explicitly because all of them
are within symbols.
Least square fitted lines for each value of $\beta$ are also shown.

\noindent {\bf Figure \ 2} \\
Monte Carlo results on the normalized frequencies of each
state $\mid \alpha \rangle$ in $(\ref{eq:Ar})$
obtained on an $N=4$ and $J'=1$ ladder
with $\beta=5$ and $n_t=32$, together with exactly calculated non-zero
$c_\alpha^2$ where $c_\alpha$'s are coefficients in the expansion of
the ground state $(\ref{eq:G})$.
The abscissa represents the state $\mid \alpha \rangle$ with non-zero
coefficients which are numbered from 1 to 38.
In each of twelve random number sequences we discarded first 2000
configurations and used the following 10000 configurations for the
measurement.

\noindent {\bf Figure \ 3} \\
Sum of the normalized frequencies on the states possessing the hidden
order in $(\ref{eq:Ar})$, which is plotted versus $J'$ for an $N=8$
ladder and versus $N$ for a $J'=1$ ladder.
Each datum is the extrapolated one from data for $n_t=16$, $20$, $24$
and $32$ to $n_t \rightarrow \infty$.
Data for finite $n_t$ are obtained with twelve random number sequences in
each of which we used 20000 configurations with 20000 preceding ones.
Errors for the extrapolation calculated from the statistical errors
are within symbols.

\noindent {\bf Figure \ 4} \\
Monte Carlo results for the non-zero coefficients $c_\alpha$'s
on an $N=4$ and $J'=1$ ladder with $\beta=5$ and $n_t=32$.
We calculated $c_\alpha$'s using the normalized frequencies obtained
from twelve simulations in each of which 10000 configurations are
observed after 2000 initial sweeps.
The abscissa is for the state numbers which are same to those in
Figure 2.
We also plot exact values for comparison.

\noindent {\bf Figure \ 5} \\
Sum of the normalized frequencies on the states
possessing the hidden order in $(\ref{eq:Ard})$
measured on a $J'=-1$ ladder with the Trotter number $n_t=32$,
which is plotted for $N=4$, 8, 12 and 16 as a function of $\beta$.
For each $\beta$ and $N$ we carried out twelve simulations with different
random number sequences in each of which we updated the system 20000 times
for the thermalization and then did additional 20000 updates
for the measurements.
The $r$ ratio ranges from $\sim 0.76$ to $\sim 0.01$.
The dashed line is the exact value for an $N=4$ ladder.

\end{document}